\documentclass[fleqn,usenatbib]{mnras}

\usepackage{newtxtext,newtxmath}

\usepackage[T1]{fontenc}
\usepackage{ae,aecompl}

\usepackage{physics}
\usepackage{cleveref}

\usepackage{graphicx}	
\usepackage{amsmath}	

\usepackage[dvipsnames]{xcolor}
\usepackage[normalem]{ulem}
\usepackage{subfig}
\usepackage{tablefootnote}

\graphicspath{{./figures/}}

\title[Planets with escaping atmospheres]{Caught in the Act: Core-powered Mass-loss Predictions for Observing Atmospheric Escape}

\author[A. Gupta and H.E. Schlichting]{
Akash Gupta$^{1}$\thanks{E-mail: akashgpt@ucla.edu} and
Hilke E. Schlichting$^{1,\;2,\;3}$
\\
$^{1}$Department of Earth, Planetary, and Space Sciences, University of California, Los Angeles, CA 90095, USA\\
$^{2}$Department of Physics and Astronomy, University of California, Los Angeles, CA 90095, USA\\
$^{3}$Department of Earth, Atmospheric and Planetary Sciences, Massachusetts Institute of Technology, MA 02139, USA
}

\date{Accepted XXX. Received XXX; in original form XXX}

\pubyear{2021}

\begin{document}
\label{firstpage}
\pagerange{\pageref{firstpage}--\pageref{lastpage}}
\maketitle

\begin{abstract}
Past studies have demonstrated that atmospheric escape by the core-powered mass-loss mechanism can explain a multitude of observations associated with the radius valley that separates the super-Earth and sub-Neptune planet populations. Complementing such studies, in this work, we present a shortlist of planets that could be losing their atmospheres today if their evolution is indeed primarily dictated by core-powered mass-loss. We use Bayesian inference analysis on our planet evolution and mass-loss model to estimate the posteriors of the parameters that encapsulate the current state of a given planet, given their published masses, radii and host star properties. Our models predict that the following planets could be losing their atmospheres today at a rate $\gtrsim 10^7$ g/s at 50\% confidence level: pi Men c, Kepler-60 d, Kepler-60 b, HD 86226 c, EPIC 249893012 b, Kepler-107 c, HD 219134 b, Kepler-80 e, Kepler-138 d and GJ 9827 d. As a by-product of our Bayesian inference analysis, we were also able to identify planets that most-likely harbor either secondary atmospheres abundant with high mean-molecular weight species, low-density interiors abundant with ices, or both. The planets belonging to this second category are WASP-47 e, Kepler-78 b, Kepler-10 b, CoRoT-7 b, HD 80653 b, 55 Cnc e and Kepler-36 b. While the aforementioned lists are by no means exhaustive, we believe that candidates presented here can serve as useful input for target selection for future surveys and for testing the importance of core-powered mass-loss in individual planetary systems.
\end{abstract}


\section{Introduction} \label{sec:intro}

Observational studies in the last decade have given us the opportunity to gain unprecedented insight into the origin of small exoplanets. Results from NASA's \textit{Kepler} mission have revealed that around FGK stars, close-in planets between the sizes of Earth and Neptune are not just common \citep[e.g.][]{fressin2013a,petigura2013a}, but that there are very few planets of intermediate sizes between 1.5 to 2.0 Earth radii \citep[R$_{\oplus}$; e.g.][]{owen2013a,fulton2017a}. 

This lack of intermediate sized planets or the bimodality in the size distribution of small planets, referred to as the `radius valley' or `radius gap', has since been corroborated by several studies \citep[e.g.][]{vaneylen2018a,fulton2018a,berger2018a,martinez2019a,berger2020b}. Furthermore, the radius valley has now even been observed in the \textit{K2} data \citep[e.g.][]{zink2020a} and for low mass stars \citep[e.g.][]{cloutier2020a,vaneylen2021a}. Follow-up surveys seeking masses of these small planets have further revealed that there is an overlapping bimodality in their compositions as well \citep[e.g.][]{marcy2014b,rogers2015a}. Planets smaller than $\sim$ 1.6 $R_{\oplus}$ have higher densities that are consistent with rocky, Earth-like compositions \citep[e.g.][]{dressing2015a,dorn2019a,bower2019a} whereas planets larger than $\sim$2 R$_{\oplus}$ have lower densities which suggests that these planets are Neptune-like and engulfed in H/He envelopes \citep[e.g.][]{jontof-hutter2016a}. It has thus been suggested that the radius valley is a transition regime from smaller, rocky planets, i.e. `super-Earths' to the larger planets with significant H/He envelopes, i.e. `sub-Neptunes' \citep[e.g.][]{rogers2015a}. 

A multitude of mechanisms have been proposed to explain the radius valley \citep[e.g.][]{owen2013a,ginzburg2018a,zeng2019a,lee2020a} but only a few can explain the numerous observations pertinent to it \citep[e.g.][]{fulton2018a,vaneylen2018a,owen2018a,loyd2020a,berger2020b}. Currently, the leading theory is that atmospheric mass-loss due to mechanisms such as photoevaporation and/or core-powered mass-loss leads to the observed bimodality in planet sizes and compositions \citep[e.g.][Gupta \& Schlichting 2019]{owen2013a,lopez2013a,ginzburg2018a}. Assuming that planets are typically born in protoplanetary gas disks and thus accrete H/He envelopes, photoevaporation and core-powered mass-loss studies argue that planets that subsequently lose their entire atmospheres are today's super-Earths whereas sub-Neptunes are those that survived with significant atmospheres.

Interestingly, studies have detected active atmospheric mass-loss from planets by looking for absorption signatures of different atomic and molecular species in their spectra. One of such signatures is the Lyman-alpha (Ly$\alpha$) line which probes the presence of neutral hydrogen. This spectral line (doublet) corresponds to a wavelength of 121.6 nm and is due to excitation of an electron from the n=1 orbital to n=2 orbital (a Lyman series transition), where n is the principal quantum state of an atom. \citet{vidal-madjar2003a} were the first to use this line to report the presence of an extended atmosphere around a planet - HD 209458 b - based on the fact that the transit depth they observed in Ly$\alpha$ was several times deeper than what was observed for an optical transit. Since excess absorption in Ly$\alpha$ has even been observed around smaller exoplanets such as GJ 436 b \citep[][]{kulow2014a,ehrenreich2015a,lavie2017a}, GJ 3470 b \citep[][]{bourrier2018a}, K2-18 b \citep[][]{dossantos2020a}, Kepler-444 e and f \citep[][]{bourrier2018a} and GJ 9827 b \citep[][]{carleo2021a}. On the other hand, non-detections in Ly$\alpha$ have been reported for small exoplanets such as 55 Cnc e \citep[][]{ehrenreich2012a}, HD 97658 b \citep[][]{bourrier2017a}, GJ 1132 b \citep[][]{waalkes2019a} and pi Men c \citep[][]{garcia2020a}. Similarly, studies have also looked for the H-alpha (H$\alpha$) line as a signature of extended atmospheres \citep[e.g.][]{jensen2012a}. This 656.3 nm spectral line corresponds to the first Balmer series transition in hydrogen. So far, the H$\alpha$ line has only been detected around giant planets such as HD 209458 b, HD 189733 b and WASP-33 b \citep[e.g.][]{jensen2012a,yan2018a}. Among small exoplanets, a non-detection has been reported for GJ 9827 b \citep[][]{carleo2021a}. Another excellent probe for atmospheric escape is the He 1083 nm triplet absorption feature \citep[][]{seager2000a,oklopcic2018a}. It has gained significant traction in the exoplanet community because it can overcome the primary shortcomings of the hydrogen lines such as absorption in the ISM and contamination from geocoronal emission. This spectral line is a consequence of a transition from the metastable 2$^3$S state - one of the spin states for helium electrons - to the 2$^3$P state by absorption at 1083 nm wavelength. Beginning with detection of atmospheric escape from WASP-107 b \citep[][]{spake2018a}, it has already been used to detect mass-loss from various other planets including the warm Neptune GJ 3470 b \citep[][]{ninan2020a}. To this date, however, the He 1083 nm line has not been detected around a planet less massive than GJ 4370 b. The small exoplanets for which non-detections have been reported include GJ 436 b \citep[][]{nortmann2018a}, K2-100 b \citep[][]{gaidos2020a}, GJ 1214 b \citep[][]{kasper2020a}, HD 97658 b \citep[][]{kasper2020a}, GJ 9827 d \citep[][]{kasper2020a,carleo2021a}, GJ 9827 b \citep[][]{carleo2021a} and 55 Cnc e \citep[][]{zhang2020a}. 

As mentioned previously, if a planet's transit depths near the aforementioned wavelengths are much larger than what is observed in an optical transit, indicates that the planet has an extended atmosphere surrounding it. In other words, this implies that there are a significant number of neutral hydrogen or metastable helium atoms at distances much greater than the photospheric radius of the planet. This is only plausible if the planet is losing its atmosphere. The depth of a transit primarily depends on the atmospheric profile or more specifically, the radial number density profile of the species being observed (e.g., neutral hydrogen for the Ly$\alpha$ line and metastable helium for the 1083 nm line), assuming reasonable knowledge of the cross-section of the absorbing species and the amount of stellar emission near the respective wavelengths. Inversely, if we know the transit depth of a planet, using an atmosphere profile model, we can estimate the number density profile of the species in question and thus, the mass-loss rate for the planet. A non-detection of excess absorption, however, doesn't necessarily imply that a planet is not undergoing atmospheric escape. Rather, it implies that the number density of the species being observed is below the detection limits and therefore, simply puts a constraint on the maximum mass-loss rate the planet could be undergoing. 

Furthermore, while it is certainly the absorption of high energy stellar radiation by numerous hydrogen and helium atoms in the extended atmosphere that leads to the observed larger transit depths in Ly$\alpha$, H$\alpha$ and the Helium 1083 nm line - it is not clear if this atmospheric escape from small exoplanets is driven by photoevaporation or by core-powered mass-loss. In other words, even if a planet's evolution is dominated by core-powered mass-loss, it is certainly also subject to high energy radiation from its host star and the pertinent photochemistry.

To help elucidate which process may dominate the atmospheric mass-loss from small exoplanets and to help inform further observational surveys, we identify a set of ten planets that  could be undergoing considerable atmospheric mass-loss today, if their evolution was dominated by core-powered mass-loss. To arrive at this list of candidates, we utilize the open-source code \texttt{dynesty} \citep[][]{speagle2020a} to construct a Bayesian inference model around our core-powered mass-loss planet evolution code, previously used in \citet[][]{gupta2019a,gupta2020a}. This way, we are able to estimate planets' physical properties, such as bulk core compositions, atmospheric masses and mass-loss rates, that best explain their observed radii and masses, given their age and the bolometric flux they receives from their host stars.

This paper is structured as follows: In \Cref{sec:method}, we discuss our methodology for this work. We first give a review of how planets evolve under the core-powered mass-loss model, followed by a description of our Bayesian inference model and finally, we discuss the selection criteria for our sample of planets. Subsequently, we discuss our results in \Cref{sec:results} that includes our list of planets which we recommend for follow-up observations and also discuss our results in the context of past observational and theoretical studies. We finally conclude this work in \Cref{sec:conclusion}.

\section{Methodology}\label{sec:method}
In this section, we discuss our approach for estimating which of the observed planets could be undergoing significant mass-loss today under the core-powered mass-loss mechanism. To this end, we first discuss the core-powered mass-loss model in \Cref{sec:core-powered mass-loss} and then our Bayesian inference model in \Cref{sec:bayes}. Finally, we lay out our sample selection criteria in \Cref{sec:sample}.

\subsection{Planet Evolution Model: Core-Powered Mass-Loss Mechanism} \label{sec:core-powered mass-loss}

In the early phases of planet formation, as a core forms and grows by accreting solids, the gravitational binding energy of the accreting material gets converted into thermal energy. This energy can be efficiently radiated away if the core forms after the protoplanetary gas disk has dispersed. However, if the accretion occurs in the presence of the gas disk, the core will accrete a H/He envelope from the surrounding nebula once its Bondi radius becomes larger than its physical radius. Once this envelope becomes optically thick, it will act as a `thermal blanket' for the planet because the loss of heat from the core will then be limited by the thermal diffusion across the radiative-convective boundary of the envelope \citep[e.g.][]{lee2015a,ginzburg2016a} - significantly increasing the cooling timescales of the underlying core. In other words, if a planet forms in the presence of a protoplanetary gas disk, it retains a significant fraction of its primordial energy from formation. In fact, the core temperature is likely set by the maximum temperature that permits the accretion of a H/He envelope which is roughly given by $T_c \sim G M_c \mu/k_B R_c$, where $\mu$ is the mean molecular mass of the atmosphere, $k_B$ is the Boltzmann constant, $G$ is the gravitational constant and $M_c$ and $R_c$ are the mass and radius of the planetary core, respectively. This implies typical core temperatures of 10$^4$ - 10$^5$ K for core masses ranging from the mass of Earth to Neptune.

Eventually, after the dispersal of the protoplanetary gas disk and the end of the spontaneous mass-loss phase/boil-off phase \citep{ginzburg2016a,owen2016a}, it is this primordial energy from planet formation together with the bolometric luminosity from the host star that can drive a Parker-type hydrodynamic outflow of the atmosphere \citep[e.g.][]{parker1958a} for small, close-in planets \citep{ginzburg2016a,ginzburg2018a,gupta2019a,gupta2020a}. We refer to this atmospheric mass-loss mechanism as core-powered mass-loss. 

To model the evolution of a planet undergoing core-powered mass-loss, we assume that a typical planet with mass $M_p$ and radius $R_p$ consists of a molten and isothermal `core' (mass $M_c$, radius $R_c$) that is surrounded by a H$_2$ atmosphere (mass $M_{atm}$, thickness $\Delta R$). We assume that these planets have most of their mass in their cores such that $M_p \sim M_c$ and that their cores and envelopes are thermally well coupled at the core-envelope interface. As in previous works, we adopt a two-layer model for the surrounding atmosphere with an inner convective and an outer radiative region \citep[e.g.][]{piso2014a,inamdar2015a,lee2015a}. We model the convective region as adiabatic and the radiative region as isothermal, and refer to their interface as the radiative-convective boundary ($R_{rcb}$). Since the radiative region is close to isothermal, the atmospheric density decreases exponentially with radial distance and most of the atmospheric mass is thus contained in the convective region. Therefore, we assume that the observed radius of a planet, or its photospheric radius, $R_p \sim R_{rcb}$.

Since we are only modelling the atmospheric loss after the spontaneous mass-loss/boil-off phase \citep[][]{owen2016a,ginzburg2016a}, we assume, following previous studies, $R_{rcb} \sim 2 R_c$ as initial condition.

We account for gravitational compression of the planetary cores by assuming that their mass-radius relation is given by 
\begin{equation}
M_c/M_{\oplus}=(R_c/R_\oplus)^4 (\rho_{c\ast}/\rho_\oplus)^{4/3}  \label{eq:Mc_Rc_reln} 
\end{equation}
where $\rho_{c\ast}$ is the density of a planet's core when scaled to Earth mass \citep[e.g.][]{valencia2006a,seager2007a,fortney2007a}.

To track the evolution of such a planet undergoing core-powered mass-loss, we simultaneously evolve the total energy a planet has available for cooling ($E_{cool}$) and its atmospheric mass ($M_{atm}$) as a function of time by numerically solving the following equations:
\begin{align}
E_{cool}(t+\text{d}t) &=  E_{cool}(t) - L_{rcb}(t) \text{ d}t \text{ and}\\
M_{atm}(t+\text{d}t) &= M_{atm}(t) - \dot{M}_{atm}(t) \text{ d}t.
\end{align}
Here $L_{rcb}$ is the luminosity of the planet at the radiative-convective boundary and $\dot{M}_{atm}$ is the rate at which the planet is losing atmospheric mass. The former can be expressed as
\begin{equation}\label{eq:L}
L_{rcb} = \frac{64 \pi}{3} \frac{\sigma T_{rcb}^4 R_B^{\prime}}{\kappa \rho_{rcb}},
\end{equation}
where $\kappa$, $T_{rcb}$ and $\rho_{rcb}$ are the opacity, temperature and density at the radiative-convective boundary, $\sigma$ is the Stefan-Boltzmann constant and $R_B^{\prime} \equiv \frac{\gamma - 1}{\gamma} {G M_c \mu/}{(k_B T_{rcb})},$ where $\gamma$ and $\mu$ are the adiabatic index and molecular weight of the atmosphere and $k_B$ is the Boltzmann constant. The mass-loss rate  ($\dot{M}_{atm}$) that a planet experiences is assumed to be the smaller of the
\begin{enumerate}
\item the energy-limited rate ($\dot{M}_{atm}^E$), i.e., the mass-loss rate given by assuming all the cooling luminosity goes into driving atmospheric mass-loss,
\begin{equation}\label{eq:M_loss_rate_E}
\dot{M}_{atm}^E \simeq \frac{L_{int}(t)}{g R_c},
\end{equation}
{where $g = G M_c/R_c^2$ is the acceleration due to gravity evaluated at the core-atmosphere boundary}, and

\item the Bondi-limited rate ($\dot{M}_{atm}^B$), i.e., the mass-loss rate that is dictated by the thermal velocity of the gas molecules at the Bondi radius,
\begin{equation}\label{eq:M_loss_rate_B}
\dot{M}_{atm}^B = 4\pi R_s^2 c_s \rho_{rcb} \; \text{exp}\left( -\frac{GM
_p}{c_s^2 R_{rcb}}\right),
\end{equation}
where $R_s=GM_p/2c_s^2$ is the radius where the atmospheric outflow reaches sonic velocity and $c_s = (k_B T_{\text{eq}}/\mu)^{1/2}$ is the isothermal speed of sound.

\end{enumerate}

Due to this evolution, planets ultimately either lose all their atmospheres and become essentially naked cores and constitute the population of smaller planets below the radius valley, i.e. super-Earths, or retain some of their primordial H/He atmosphere and remain part the population of larger planets above the radius valley, i.e. sub-Neptunes (see, for example, Figure 3 from \citet{gupta2020a} and for further details \citet{ginzburg2018a} and \citet{gupta2019a}).

\subsection{Bayesian Inference Model} \label{sec:bayes}
We use Bayesian inference analysis to estimate if a planet, given its mass, radius and its host star's properties, could be undergoing significant mass-loss today. Under the Bayesian framework, Bayes' theorem is used to estimate the posterior distribution $P({\bf\Phi}|\textbf{D}, M)$ of a set of parameters ${\bf \Phi}$ for a model $M$, given some data $\textbf{D}$ and our prior knowledge of $\bf \Phi$, i.e.
\begin{align}
    P({\bf\Phi}|\textbf{D}, M) &= \frac{P(\textbf{D}|{\bf\Phi}, M) \times P({\bf\Phi}|M)}{P(\textbf{D}|M)},
\end{align}
where $P(\textbf{D}|{\bf\Phi}, M)$ quantifies the likelihood of observing the data $\bf D$ given the model $M$ and the parameters $\bf \Phi$, i.e. it's the likelihood distribution. $P({\bf\Phi}|M)$ is known as the prior distribution and signifies our prior knowledge of the parameters $\bf \Phi$ that we input in the model $M$. $P(\textbf{D}|M)$, which can be expressed as
\begin{align}
    P(\textbf{D}| M) &= \int_{\forall {\bf\Phi}} {P(\textbf{D}|{\bf\Phi}, M) \times P({\bf\Phi}|M)}\;\text{d}{\bf\Phi},
\end{align}
embodies our confidence in the model $M$ given the observations.

In the context of this work,
\begin{enumerate}
    \item $M$ is our planet evolution model based on the core-powered mass-loss mechanism as discussed in \Cref{sec:core-powered mass-loss}, 
    \item $\textbf{D} = \{... , \text{D}_i, ... \}$ corresponds to the observed properties of a planet and its host star, namely, the observed radius ($R_p$), mass ($M_p$) and insolation flux ($S_p$) of a given planet and the estimated age of its host star ($\tau_{\ast}$), and 
    \item ${\bf \Phi} = \{... , \Phi_i, ... \}$ includes all the input parameters for our model \textit{M} such as a planet's primordial atmosphere mass fraction ($f_{\text{initial}}$), bulk density of the core scaled to an Earth mass ($\rho_{c\ast}$), $M_p$, $R_p$, $S_p$ and $\tau_\ast$.
\end{enumerate}

For any observed property D$_i$, we assume that its uncertainties and thus its likelihood are Gaussian in nature, centered at $\mu_i$ (= the nominal or measured value) with a standard deviation $\sigma_i$ (= (upper error limit + lower error limit)/2). The total likelihood, given $\bf \Phi$, is then the product of the individual likelihoods corresponding to each D$_i$. For all parameters $\in {\bf\Phi}$, except for $\rho_{c\ast}$, we assume uniform priors. For $\rho_{c\ast}$, we assume a Gaussian distribution centered at $\mu_{\rho_{c\ast}}=$ 5 g/cm$^{3}$ with a $\sigma_{\rho_{c\ast}}=$ 1 g/cm$^{3}$. This is motivated by recent mass and radius measurements of rocky planets \citep[e.g.][]{dressing2015a,bower2019a,dorn2019a}, geochemical studies of white-dwarf pollution \citep[e.g.][]{doyle2019a} and previous planet population studies involving evolution under core-powered mass-loss \citep{gupta2020a}. In addition, this assumption is also consistent with the results of photoevaporation studies \citep[e.g.][]{rogersj2020a}. Nevertheless, we experiment with other values for \{$\mu_{\rho_{c\ast}}, \sigma_{\rho_{c\ast}}$\} too and discuss this later in \Cref{sec:losing-mass}.

To compute the resulting $P({\bf\Phi}|\textbf{D}, M)$, we use the open-source code \texttt{dynesty} \citep{speagle2020a} which uses a Dynamic Nested Sampling algorithm to estimate posteriors and evidences \citep{skilling2004a,skilling2006a,higson2019a,feroz2009a}. Therefore, given the planet and host star observations $\textbf{D}$ and our evolution model $M$, \texttt{dynesty} allows us use the Bayes' theorem to estimate the distribution of the most likely values that planet parameters, ${\bf\Phi}$ = $\{f_{\text{initial}}, \:\rho_{c\ast},\: ...\}$, can have and their co-variances (see for e.g. \Cref{fig:pi Men c}). Knowledge of planet parameters such as the primordial atmospheric mass-fraction ($f_{\text{initial}}$) and bulk composition ($\rho_{c\ast}$) allows us to then determine the planet's current atmospheric mass-fraction ($f_{final}$) and thus its current atmospheric mass-loss rate ($\dot{M}_{\text{atm}}$).

\begin{figure*}
\centering
\includegraphics[width=1.0\textwidth,trim=290 450 380 365,clip]{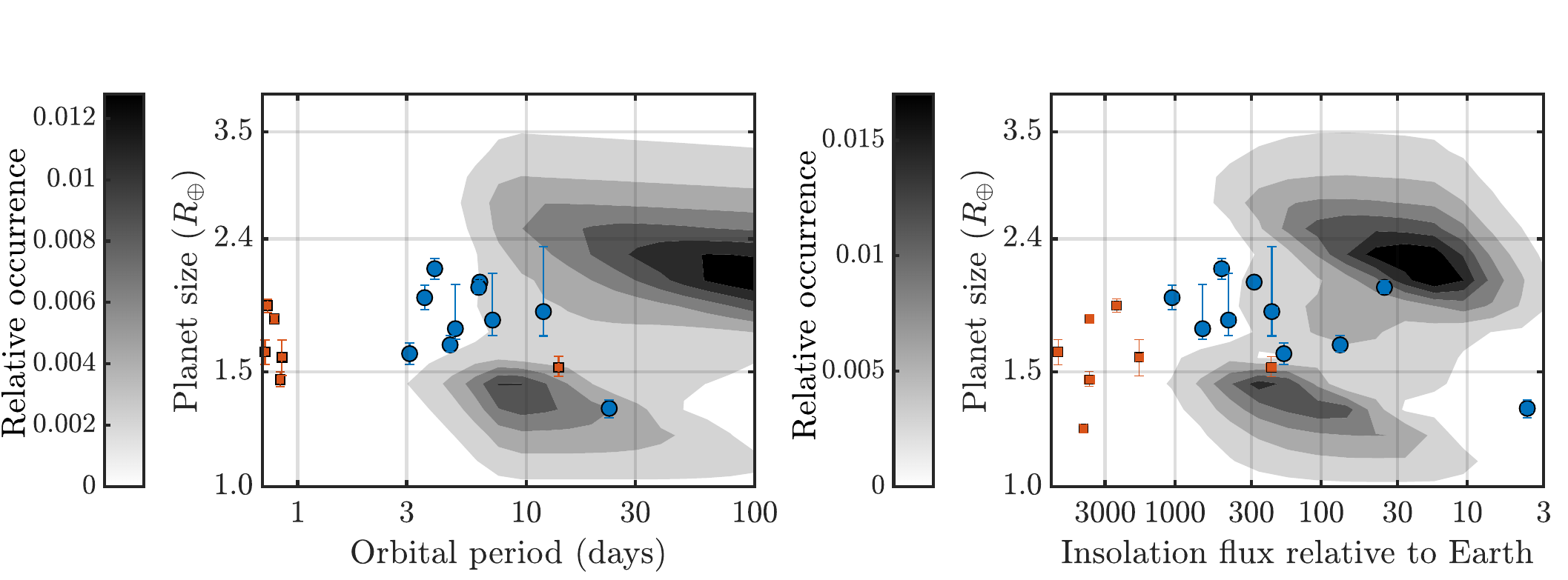} 
\caption{Location of planets for which we predict that they could be undergoing considerable atmospheric mass-loss today ({\color{NavyBlue}${\bullet}$}) in the planet size-orbital period space (left) and the planet size-insolation flux space (right). In addition, we also show planets which we identified to likely have interiors with a significant ice-fraction, or a secondary atmosphere, or both ({{\color{Bittersweet}$\blacksquare$}}). The planet size occurrence contours are from a previous core-powered mass-loss study \citep[][]{gupta2020a} which reproduced the radius valley observed around FGK stars \citep[][]{fulton2018a}. These plots show that the planets that could be losing their atmospheres today (shown in blue) lie in the radius valley.}
\label{fig:planet_locns}
\end{figure*}

\begin{table*}
\centering
{
\renewcommand{\arraystretch}{2}
\begin{tabular}{c  c  c  c  c  c  c  c  c} 
\hline
Planet & $\tau_\ast$ (Gyr) & $M_{\ast}$ ($M_{\odot}$) & $S_{p}$ ($S_{\oplus}$) & $R_p$ ($R_{\oplus}$) & $M_p$ ($M_{\oplus}$) & References \\ 
\hline
pi Men c & $5.20^{+1.10}_{-1.10}$ & $1.02_{0.03}^{0.03}$ & $282.66_{+12.42}^{-12.42}$ & $2.06^{+0.03}_{-0.03}$ & $4.52^{+0.81}_{-0.81}$  & \citet{gandolfi2018a}\\ 

Kepler-60 d & $4.80^{+2.02}_{-1.37}$ & $1.13^{+0.07}_{-0.08}$ & $217.11_{+16.64}^{-18.63}$ & $1.86^{+0.48}_{-0.15}$ & $3.90^{+0.70}_{-0.60}  \phantom{}^{\dagger}$  & \citet{berger2020a,berger2020b}\\

Kepler-60 b & $4.80^{+2.02}_{-1.37}$ & $1.13^{+0.07}_{-0.08}$ & $429.50_{+32.93}^{-36.87}$ & $1.80^{+0.32}_{-0.10}$ & $3.70^{+0.60}_{-0.60} \phantom{}^{\dagger}$ & \citet{berger2020a,berger2020b}\\

HD 86226 c & $4.60^{+3.70}_{-2.70}$ & $1.02^{+0.06}_{-0.07}$ & $481.92_{+42.20}^{-42.20}$ & $2.16^{+0.08}_{-0.08}$ & $7.25^{+1.19}_{-1.12}$  & \citet{teske2020a}\\ 

EPIC 249893012 b & $9.00^{+0.50}_{-0.60}$ & $1.05^{+0.05}_{-0.05}$ & $1044.23_{+88.10}^{-88.10}$ & $1.95^{+0.09}_{-0.08}$ & $8.75^{+1.09}_{-1.08}$ & \citet{hidalgo2020a}\\ 

Kepler-107 c & $3.23^{+1.27}_{-0.94}$ & $1.25^{+0.05}_{-0.06}$ & $645.38_{+48.35}^{-46.28}$ & $1.75^{+0.30}_{-0.07}$ & $9.39^{+1.77}_{-1.77} {\phantom{}^{\ast}}$ & \citet{berger2020a,berger2020b}\\ 

HD 219134 b & $11.00^{+2.20}_{-2.20}$ & $0.81^{+0.03}_{-0.03}$ & $179.59_{+5.51}^{-5.51}$ & $1.60^{+0.06}_{-0.06}$ & $4.74^{+0.19}_{-0.19}$ & \citet{gillon2017a}\\ 

Kepler-80 e & $10.74^{+6.09}_{-6.70}$ & $0.71^{+0.03}_{-0.03}$ & $73.96_{+5.36}^{-4.86}$ & $1.65^{+0.05}_{-0.04}$ & $4.13^{+0.81}_{-0.95} {\phantom{}^{\P}}$ & \citet{berger2020a,berger2020b}\\ 

Kepler-138 d & $10.81^{+6.16}_{-6.96}$ & $0.54^{+0.01}_{-0.01}$ & $3.88_{+0.34}^{-0.32}$ & $1.32^{+0.04}_{-0.04}$ & $1.17^{+0.30}_{-0.30} {\phantom{}^{\S}}$ & \citet{berger2020a,berger2020b}\\ 

GJ 9827 d & $10.00^{+2.00}_{-5.00}$ & $0.61^{+0.02}_{-0.02}$ & $36.83_{+1.86}^{-1.86}$ & $2.02^{+0.05}_{-0.04}$ & $4.04^{+0.82}_{-0.84}$ & \citet{rice2019a}\\

\hline

\end{tabular}
\caption{List of planets that could be undergoing mass-loss today and their observed parameters (host star's age ($\tau_\ast$) and mass (M$_\ast$), planet's insolation flux (S$_p$), radius (R$_p$), mass (M$_p$) and density when scaled to an Earth mass ($\rho_{p\ast}$)). Values are reported with 1 $\sigma$ errors. The last column lists the references from where all the planet and host-star properties are taken from, unless noted otherwise below. Mass estimates for the Kepler planets were taken from $^{\dagger}$\citet{hadden2017a}, $^{\ast}$\citet{bonomo2019a}, $^{\P}$\citet{macdonald2016a} and $^{\S}$\citet{almenara2018a}.
}
\label{table:sim_char}
}
\end{table*}

\subsection{Sample Selection} \label{sec:sample}

On August 12, 2020, we collected all the planets from the NASA Exoplanet Archive that have masses below $20 M_\oplus$ at 2$\sigma$ confidence level and an age estimate for their host stars. We then further restricted the sample to planets for which mass measurements have uncertainties lower than 30$\%$. 

Furthermore, we removed planets with host star masses below 0.5 $M_\odot$, because to accurately model the evolution of planets around such low-mass host stars one needs to take into account their Gyr-long pre-main-sequence evolution and the related bolometric luminosity changes \citep[e.g.][]{ramirez2014a}, which is beyond the scope of this paper. After all these cuts, we were left with a sample of 78 planets.

Most of the Kepler planets from this sample are also present in a recently published  \textit{Gaia}-\textit{Kepler} Stellar Catalogue by \citet{berger2020a,berger2020b} that has  homogeneously-derived stellar and planetary properties. For the common \textit{Kepler} planets, we thus updated all the stellar and planetary properties, except planet mass, from \citet{berger2020a,berger2020b}.

\section{Results} \label{sec:results}

Below we identify planetary candidates for which we predict that their atmospheric loss could be observable today if their evolution has been dominated by core-powered mass-loss. In addition, we provide a list of planetary candidates that likely either have secondary atmospheres or interiors with significant ice-fraction by mass, or both.

In our numerical simulations, we assume that a planet has become a super-Earth when $f_\text{final} = M_{atm}/M_c < 10^{-10}$ or $\Delta R/R_c < 10^{-3}$.\footnote{{Note that the typical uncertainties in planet radii measurements are much larger at $\sim 5\%$ \citep[]{fulton2018a}}} For these planets, we set $f_{final}=0$, $\Delta R/R_c = 0$ and $\dot{M}_{atm} = 10^{-10}$ g/s. Giving super-Earths a tiny, but none-zero, mass-loss rate allows us to include it in the mass-loss rate posterior plots which have logarithmic scaling. We confirmed that our results do not depend on the exact choice of our chosen cut-offs.

\begin{table*}
\centering
{
\renewcommand{\arraystretch}{2}
\begin{tabular}{c  c  c  c} 
\hline
Planet &  \multicolumn{3}{c}{ $\dot{\text{M}}_{atm}$ [log$_{10}$(g/s)] } \\
 & $\mu_{1/2}$ (50$^{th}$-percentile) & [$\mu_{1/2}-\sigma$, $\mu_{1/2}+\sigma$] &  [$\mu_{1/2}-2\sigma$, $\mu_{1/2}+2\sigma$] \\
\hline
pi Men c & $9.40$ & [8.98, 9.69] & [7.96, 9.90]\\ 

Kepler-60 d & $8.85$ & [-10, 9.77] & [-10, 10.18]\\

Kepler-60 b  & $8.83$ & [-10, 9.52] & [-10, 9.96]\\

HD 86226 c  & $8.21$ & [7.14, 9.20] & [4.76, 9.80]\\ 

EPIC 249893012 b  & $8.08$ & [5.51, 8.54] & [-10, 9.02]\\ 

Kepler-107 c & $7.69$ & [-10, 8.45] & [-10, 9.13]\\ 

HD 219134 b &  $7.59$ & [-10, 7.94] & [-10, 8.22]\\ 

Kepler-80 e &  $7.44$ & [5.99, 8.15] & [-10, 8.83]\\ 

Kepler-138 d & $7.24$ & [6.21, 8.21] & [-10, 8.71]\\ 

GJ 9827 d & $6.72$ & [3.33, 8.23] & [-0.25, 9.57]\\

\hline
\end{tabular}
\caption{Core-powered mass loss predictions for planets undergoing mass-loss today. The three columns of mass-loss rates correspond to the median (50$^{th}$-percentile, $\mu_{1/2}$) and the range spanning 1 $\sigma$ and 2 $\sigma$ around $\mu_{1/2}$, respectively.}
\label{table:losing_atm}
}
\end{table*}

\subsection{Core-powered Mass-loss Predictions for Atmospheric escape observations}\label{sec:losing-mass}

We run our hybrid planet evolution-Bayesian inference model, for each of the 78 planets in our sample. Our planet evolution model neglects an atmosphere's self-gravity which is an appropriate assumption only for atmosphere mass-fractions $\lesssim 0.2$ \citep[e.g.][]{piso2014a,ginzburg2016a}. We therefore exclude from our list all 34 planets for which our evolution model predicts median initial atmosphere mass fractions $> 0.2$. Since we are trying to determine which planets might be undergoing significant mass-loss today, we also exclude all planets for which we predict current median mass-loss rates of  $\lesssim 10^7$ g/s. Finally, we decided not to include Kepler-99 b because, even though it satisfies the above two criteria, it is very sensitive to the prior we choose for $\rho_{c\ast}$. We are thus left with a list of ten planets consisting of: pi Men c, Kepler-60 d, Kepler-60 b, HD 86226 c, EPIC 249893012 b, Kepler-107 c, HD 219134 b, Kepler-80 e, Kepler-138 d, and GJ 9827 d. We show these planets as blue circles in planet size-orbital period space and planet size-insolation flux space in \Cref{fig:planet_locns}. In addition, we present the relevant planet and host star parameters in \Cref{table:sim_char} and our estimated mass-loss rates in \Cref{table:losing_atm}. We now discuss our results for each listed planet in detail.\\

\begin{figure*}
\centering
\includegraphics[width=1.0\textwidth,trim=0 0 0 0,clip]{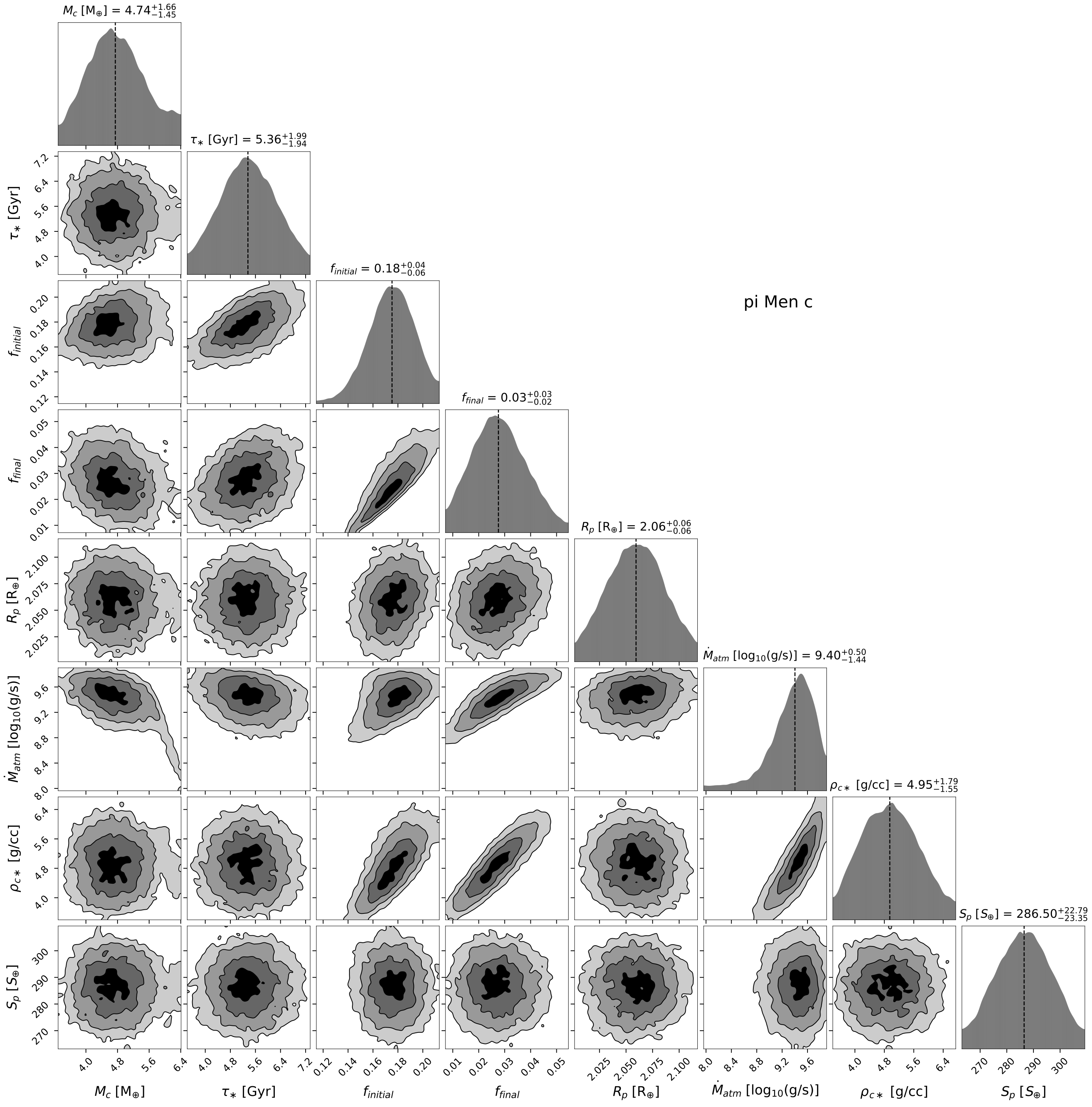} 

\caption{Corner plot for pi Men c generated using our hybrid planet evolution-Bayesian inference model. This figure shows one-dimensional posterior probability distributions of different parameters that characterize the evolution of pi Men c and their covariances. The parameters shown are planet mass (M$_p$), age ($\tau_{\ast}$), initial and final atmosphere mass fractions (f$_{initial}$ and f$_{final}$), planet radius (R$_p$), mass-loss rate ($\dot{M}_{atm}$), density of the planet's core when scaled to Earth mass ($\rho_{c\ast}$) and insolation flux received by the planet scaled to Earth (S$_p$). All parameters are reported with 2 $\sigma$ errors and vertical dashed lines denote the median values. The mass-loss rate posterior distributions show that this planet has a median mass-loss rate of $\sim 2.5 \times  10^{9}$ g/s. Our analysis thus places pi Men c in the list of planets that could be losing their atmospheres today at a considerable rate.}
\label{fig:pi Men c}
\end{figure*}

\textbf{pi Men c}: As evident from \Cref{table:losing_atm}, pi Men c is one of the top planetary candidates that could be undergoing mass-loss today. \Cref{fig:pi Men c} is a corner plot for pi Men c generated using our hybrid planet evolution-Bayesian inference model described in \Cref{sec:method}. It shows posterior distributions of the different parameters that characterize the evolution of this planet given our model. Specifically, the sub-plots in \Cref{fig:pi Men c} show the covariance and one-dimensional posterior distributions of planet mass (M$_p$), age ($\tau_{\ast}$), initial and final atmosphere mass fractions (f$_{initial}$ and f$_{final}$), planet radius (R$_p$), mass-loss rate ($\dot{M}_{atm}$), density of the planet's core when scaled to Earth mass ($\rho_{c\ast}$) and insolation flux with respect to Earth (S$_p$).

Situated only 18 parsecs away, pi Men c has a bright G0 V host star with an apparent magnitude (V) = 5.65 mag which makes it especially suitable for follow up observations \citep[][]{gandolfi2018a}. This planet has thus already been a subject of several studies and proposals. \citet{garcia2020a} recently reported non-detection of Ly$\alpha$ absorption for this planet. These authors also used a photoevaporation model to compute mass-loss rates for pi Men c and found estimates of (4 $\times$ $10^9-10^{10}$) g/s \citep[see also,][]{shaikhislamov2020a}. This is a higher mass-loss rate than estimated for GJ 436b which does show Ly$\alpha$ absorption \citep[e.g.][]{kulow2014a}. \citet{garcia2020a} thus argued that the atmosphere is likely not hydrogen-dominated and should rather have substantial amounts of heavier molecules like CO$_2$ and H$_2$O. An atmosphere that is not hydrogen-dominated, however, is unlikely to be consistent with pi Men c's low density.

As shown in \Cref{fig:pi Men c}, our core-powered mass-loss model predicts mass-loss rates of 9 $\times 10^7$ g/s to 8 $\times 10^9$ at 95\% confidence level. Since our mass-loss rates span a lower range than the predictions from photo-evaporation models, they may offer an explanation for the non-detection of Ly$\alpha$ absorption. In addition, our results imply that current observations do not necessarily preclude the possibility of pi Men c having a hydrogen-dominated atmosphere.

\begin{figure*}
\centering
\includegraphics[width=1.0\textwidth,trim=0 0 0 0,clip]{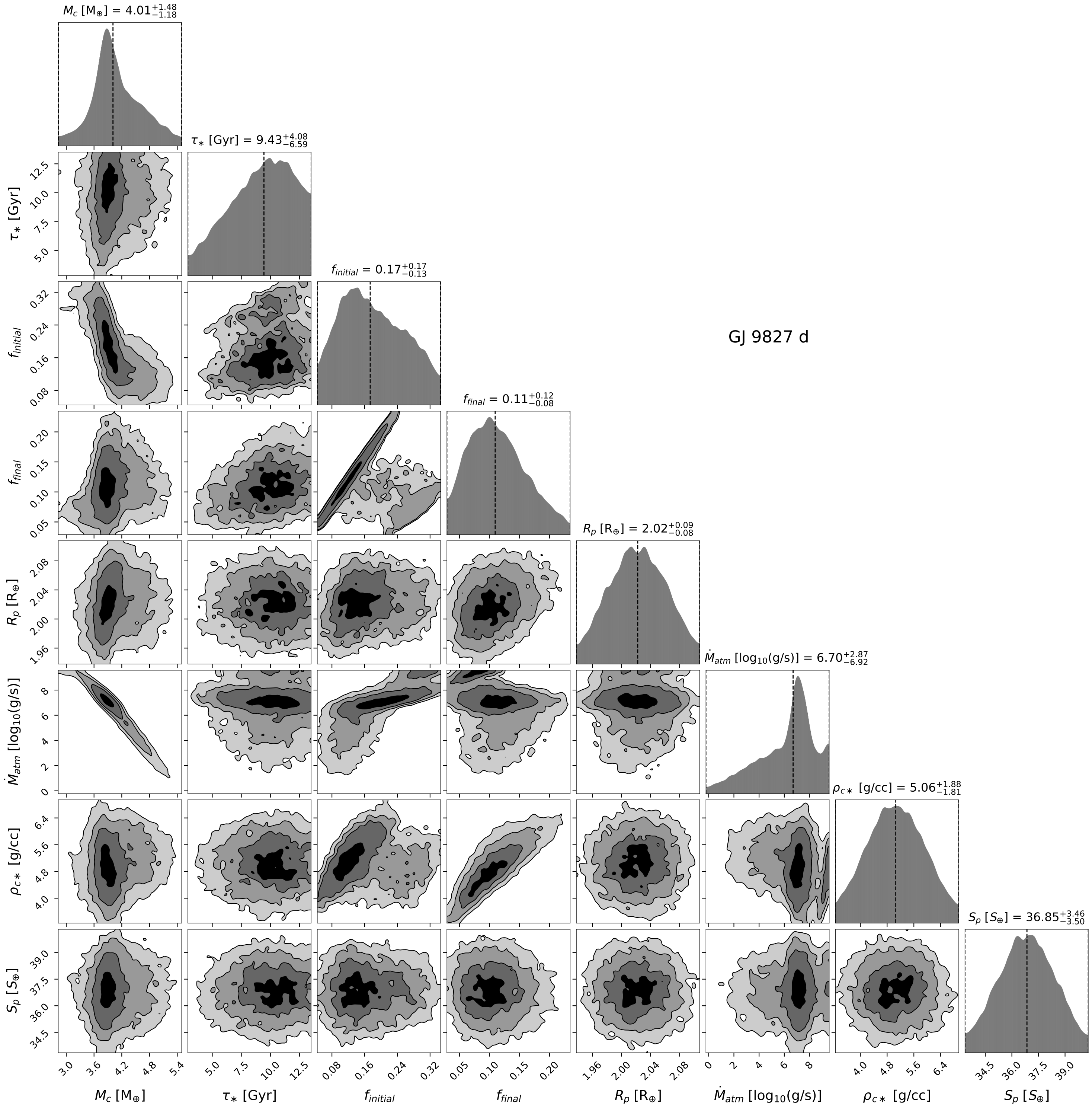} 

\caption{Corner plot for GJ 9827 d generated using our hybrid planet evolution-Bayesian inference model; similar to \Cref{fig:pi Men c}. All parameters are reported with 2 $\sigma$ errors and vertical dashed lines denote the median values. The mass-loss rate posterior shows that the 95\% confidence interval for mass-loss rate is $\sim$[0, $4 \times 10^{9}$] g/s with a median at $\sim 5 \times 10^{6}$ g/s. Our analysis thus places GJ 9827 d among planets that make interesting targets for observing atmospheric mass-loss. The covariance plot of mass-loss rate and planet mass shows how sensitive the atmospheric mass-loss rate is to the uncertainties in the planet mass measurements. Mass-loss rate estimates for GJ 9827 d vary by $\sim$ 10 orders of magnitude for planet masses within 2$\sigma$ of the median.}
\label{fig:GJ 9827 d}
\end{figure*}

\begin{figure*}
\centering
\includegraphics[width=1.0\textwidth,trim=0 0 0 0,clip]{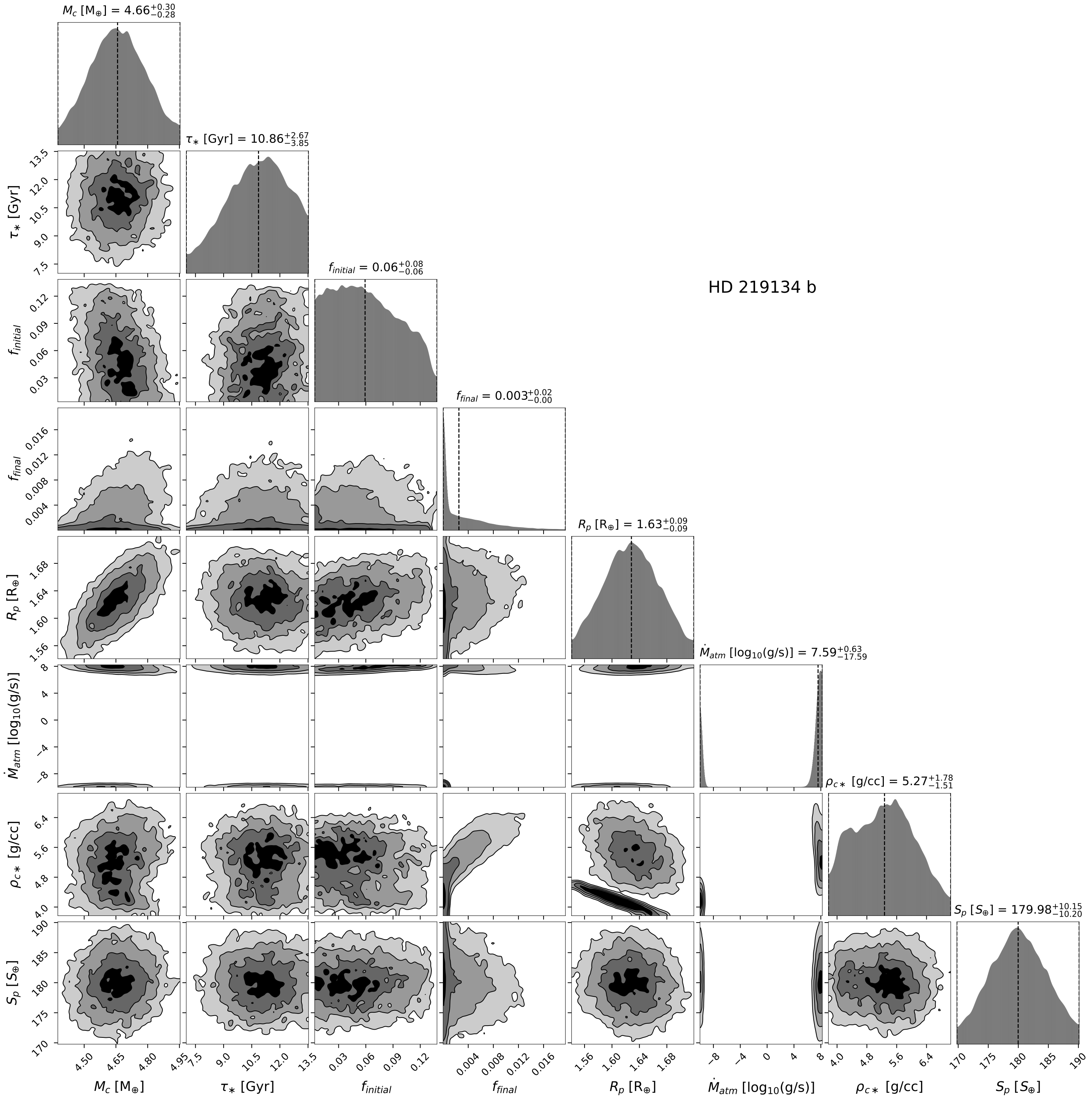} 

\caption{Corner plot for HD 219134 b generated using our hybrid planet evolution-Bayesian inference model; similar to \Cref{fig:pi Men c,fig:GJ 9827 d}. All parameters are reported with 2$\sigma$ errors and vertical dashed lines denote the median values. As with pi Men c and GJ 9827 d, our analysis places HD 219134 b in the category of planets that could be undergoing considerable atmospheric escape today but, unlike these planets, HD 219134 b has a double-peaked mass-loss posterior distribution with local maxima at $\sim$ 0 g/s and $\gtrsim 10^7$ g/s. The peak in the mass-loss rate at $\gtrsim 10^7$ g/s is consistent with GJ 9827 d having an Earth-like core with a significant hydrogen envelope. On the other hand, the peak at $\sim$ 0 g/s signifies that there is a $\sim$21\% likelihood that the planet is not losing any atmosphere and rather has a secondary, high-mean molecular weight atmosphere, a low density, icy interior, or both (see  $\rho_{c\ast}$ posteriors which show two peaks roughly centered at 4 and 5.5 g/cm$^3$).}
\label{fig:HD 219134 b}
\end{figure*}

\begin{figure*}
\centering
\includegraphics[width=1.0\textwidth,trim=0 0 0 0,clip]{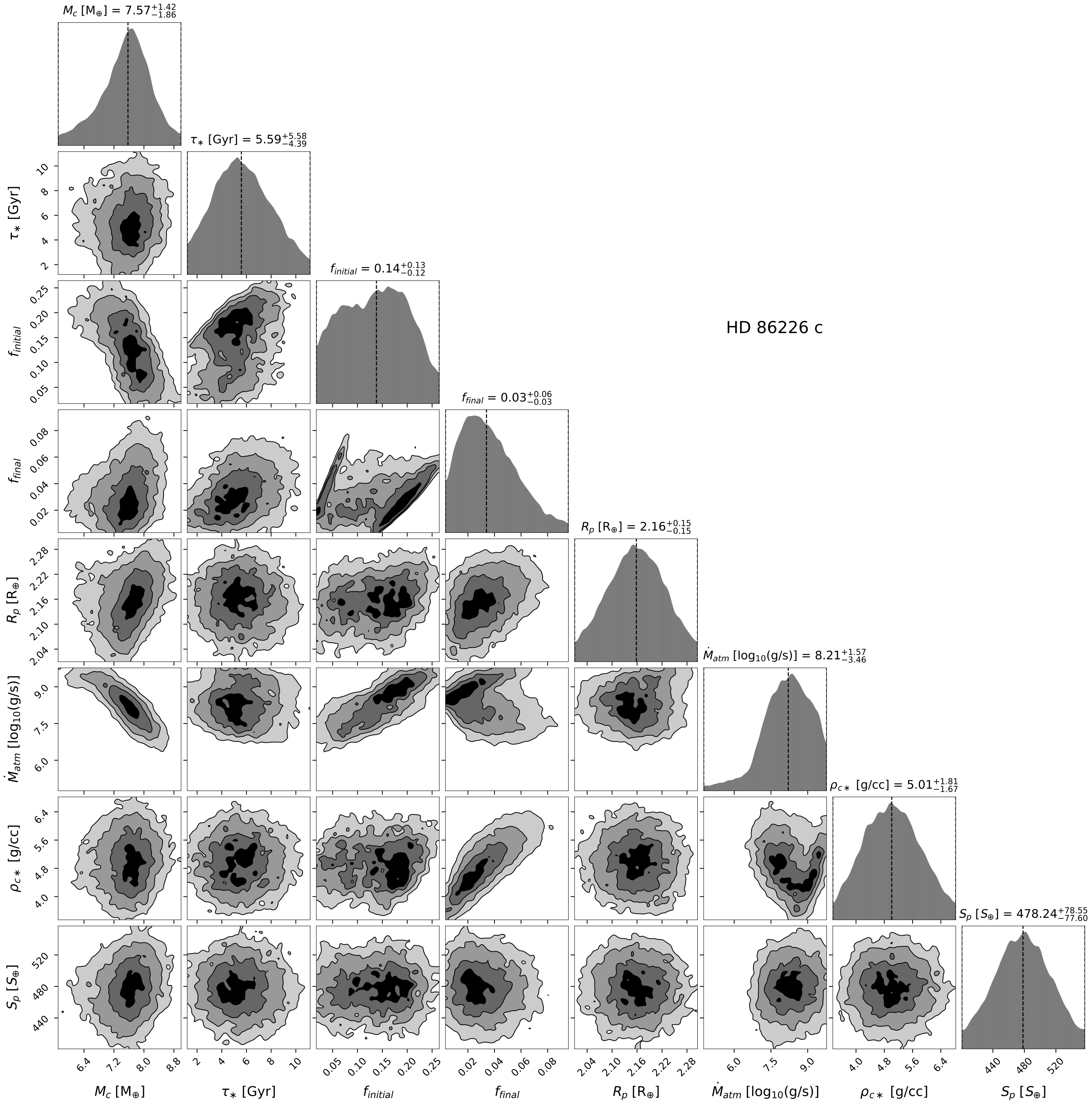} 

\caption{Corner plot for HD 86226 c generated using our hybrid planet evolution-Bayesian inference model; similar to \Cref{fig:pi Men c,fig:GJ 9827 d,fig:HD 219134 b}. All parameters are reported with 2 $\sigma$ errors and vertical dashed lines denote the median values. The mass-loss rate posterior shows that the planet has a median mass-loss rate of $\sim 1.6 \times 10^{8}$ g/s and thus, HD 86226 c is another excellent candidate for observing atmospheric mass-loss today.}
\label{fig:HD 86226 c}
\end{figure*}

\begin{figure*}
\centering

\includegraphics[width=0.85\textwidth,trim=0 0 0 0,clip]{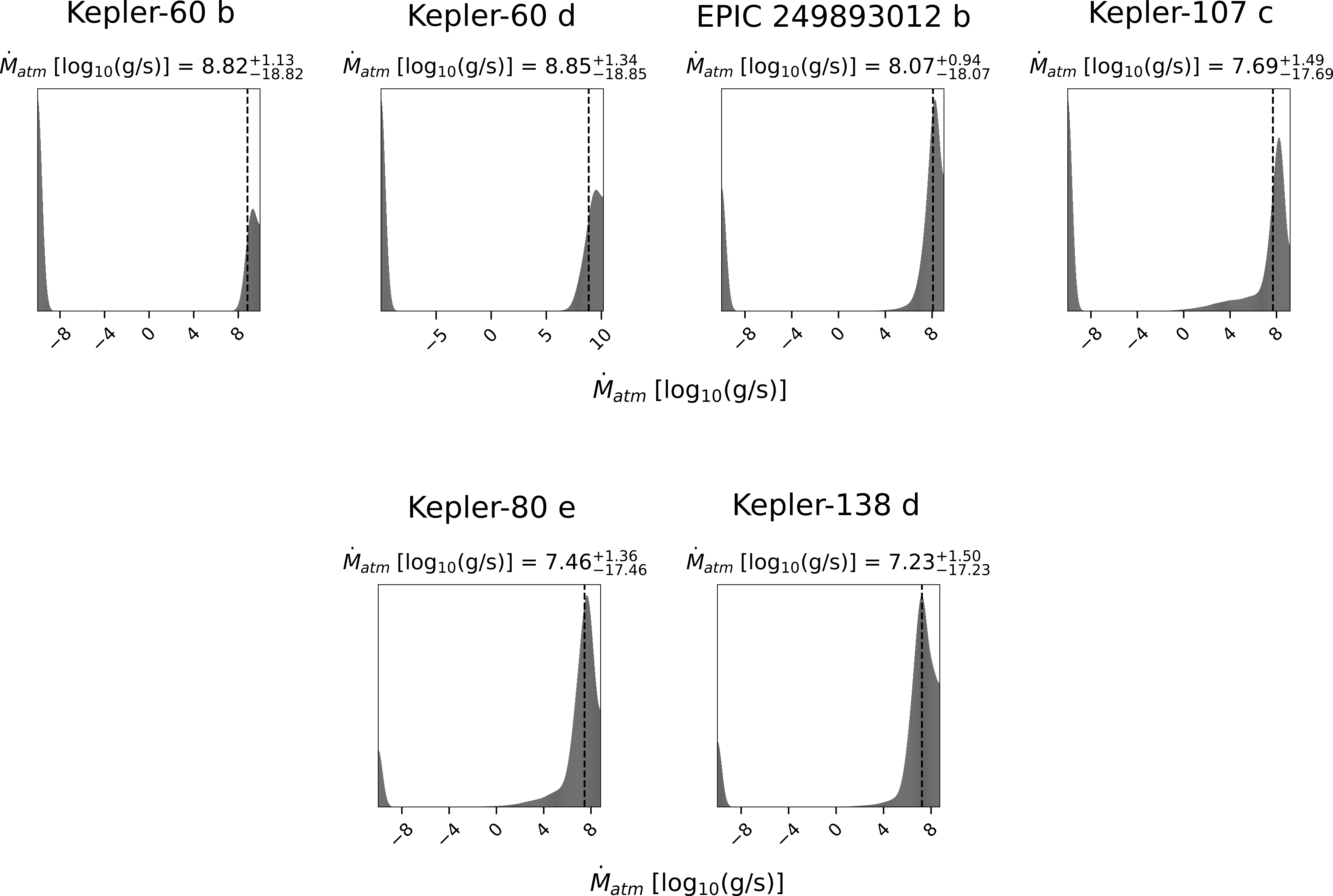}

\caption{Mass-loss rate posterior distributions for planets that, in addition to pi Men c, GJ 9827 d, HD 219134 b and HD 86226 c, could be experiencing considerable atmospheric escape at this moment: $\gtrsim 10^7$ g/s at 50$^{th}$-percentile. Mass-loss rates are reported with 2 $\sigma$ errors and vertical dashed lines denote the median values. The two-peaked nature of mass-loss rate distributions for these planets suggests that, like HD 219134 b, there is a small likelihood that these planets might not be undergoing atmospheric escape and instead, have an atmosphere abundant with heavy molecular weight species, interiors with a significant amount of ices, or both. We find that the probability of this alternate scenario is $\sim32\%$ for Kepler-60 d, $\sim42\%$ for Kepler-60 b, $\sim1\%$ for EPIC 249893012 b, $\sim14\%$ for Kepler-107 c, $\sim5\%$ for Kepler-80 e and $\sim6\%$ for Kepler-138 d.}
\label{fig:losing_atm}
\end{figure*}

Similar to \Cref{fig:pi Men c}, \Cref{fig:GJ 9827 d,fig:HD 219134 b,fig:HD 86226 c} show the corner plots GJ 9827 d, HD 219134 b and HD 86226 c. In addition, \Cref{fig:losing_atm} shows the mass-loss rate posteriors for all the other planets mentioned in \Cref{table:losing_atm}. Among the listed planets, apart from pi Men c, atmospheric observations have only been published for GJ 9827 d. There are, however, multiple theoretical studies that have investigated the existence and nature of atmospheres on GJ 9827 d and HD 219134 b.\\

\textbf{GJ 9827 d}: Part of a three planet system situated 30 parsecs away and orbiting a K6 V star with V = 10.4 mag, GJ 9827 d is another suitable target for atmospheric mass-loss observations \citep{niraula2017a,rodriguez2018a}. Recent studies by \citet{kasper2020a} and \citet[][]{carleo2021a} have reported non-detection of helium in the upper atmospheres of GJ 9827 d. \citet[][]{carleo2021a} also reported a non-detection in H$\alpha$. Using a MESA-based planet evolution model that includes mass-loss via photoevaporation \citep[e.g.][]{malsky2020a}, \citet[][]{kasper2020a} found that this non-detection is inconsistent with predicted mass-loss rate estimates of the order of $\sim 10^9$ g/s from photoevaporation. Similarly, \citet[][]{carleo2021a} predicted even higher mass-loss rates from photoevaporation of about $5.1\times10^{10}$ g/s. 

In contrast, assuming planet evolution and atmospheric loss is dominated by core-powered mass-loss, we predict that the mass-loss rate could be much lower. Specifically, we find a median mass-loss rate of $5 \times 10^6$ g/s with a 95\% confidence interval of ranging from close to zero to  $\sim 4 \times 10^9$ g/s (for details, see \Cref{table:losing_atm} or \Cref{fig:GJ 9827 d}). The non-detection of the He 1083 nm triplet feature or H$\alpha$ absorption is therefore consistent with predictions from our core-powered mass-loss model.

It is important to note that, as shown in \Cref{fig:pi Men c,fig:GJ 9827 d,fig:HD 219134 b,fig:HD 86226 c}, atmospheric mass-loss rates are especially sensitive to the uncertainties in planet mass measurements. This is particularly apparent in the covariance plot showing the correlation between mass-loss rate and planet mass for GJ 9827 d. In this specific example, mass-loss rates vary by $\sim$ 10 orders-of-magnitude for planet masses within $2 \sigma$ of the median (see \Cref{fig:GJ 9827 d}). This extreme sensitivity of the mass-loss rate on planet mass is a consequence of the fact that, given the mass, radius, age and insolation flux received by GJ 9827 d and others, the mass-loss rates are in the Bondi-limited regime (see \Cref{sec:core-powered mass-loss}). In this regime, the mass-loss rate has an exponential dependence on planet mass (see \Cref{eq:M_loss_rate_B}) and scales as
\begin{align}
    \dot{M}_{atm} &\propto \text{exp}\left(-\frac{G M_p}{c_s^2 R_{p}}\right)
   \propto \text{exp}\left(-\zeta {M_c}{S_{p}^{-1/4} R_p^{-1}}\right),
\end{align}
where $\zeta$ is a constant. The last expression shows that the mass-loss rate, for a given insolation flux and planet size, decreases exponentially with increasing planet mass and thus explains the trends in the relevant covariance plots in \Cref{fig:pi Men c,fig:GJ 9827 d,fig:HD 219134 b,fig:HD 86226 c}. Therefore, one of the best ways to improve mass-loss estimates for any of the prime candidate planets, is to get better constraints on their masses.\\

\textbf{HD 219134 b}: This planet is part of a six planet system orbiting a bright (V = 5.57 mag) K3 V star that is six parsecs away \citep[][]{vogt2015a,motalebi2015a}. With host star characteristics suitable for observations \citep[e.g.][]{gillon2017a} and a high likelihood for undergoing mass-loss, according to our analysis, this planet is another excellent target for atmospheric mass-loss observations with a predicted mass-loss rate of  $\gtrsim 10^8$ g/s at 50$^{th}$-percentile.

Interestingly, our analysis finds that mass-loss rate posteriors for HD 219134 b, and a number of other planets, are double-peaked, i.e. they peak at $\sim 0$ g/s and $\gtrsim 10^7$ g/s. While for pi Men c we estimate a mass-loss rate that is $\gtrsim$ 10$^8$ g/s at 95\% confidence level, the same can't be said for most of the other candidates; see \Cref{fig:HD 219134 b,fig:losing_atm}. Even though all the candidates are robust to our choice of $\rho_{c\ast}$, as we later discuss, the double-peaked distribution signifies that, although much less probable, there is another physically motivated solution for the composition of some of the listed planets, with bulk core densities significantly lower than Earth and/or a higher mean-molecular weight envelope (see \Cref{fig:HD 219134 b,fig:losing_atm}). That this second solution corresponds to low core densities in our model can be seen in the $\rho_{c\ast}$ posteriors in \Cref{fig:HD 219134 b}, which show peaks around 4 and 5.5 g/cm$^3$. Planets that exhibit this behavior, like HD 219134 b, are typically near the lower edge of the radius valley.

Our results therefore indicate that HD 219134 b's current state is consistent with both - having a hydrogen atmosphere and with having a low density interior (e.g. significant ice-fraction by mass or a secondary atmosphere, or both). Despite these two distinct solutions, we find that it is much more likely that HD 219134 b has a hydrogen envelope and that it therefore should be undergoing atmospheric  mass-loss today with predicted rates of $\gtrsim 10^8$ g/s at 50$^{th}$-percentile

Past theoretical studies that examine mass-loss due to photoevaporation have concluded that HD 219134 b is most likely not hydrogen-dominated because it should have lost its atmosphere by this day and therefore, should rather have a secondary atmosphere, an icy interior or a magma ocean \citep{dorn2018a,kubyshkina2018a,ligi2019a}. While no published study has yet attempted observing atmospheric mass-loss from HD 219134 b, it appears to be an especially  interesting target because photoevaporation and core-powered mass-loss studies make opposite predictions for this planet. Specifically, detecting atmospheric loss from HD 219134 b would provide direct evidence for core-powered mass-loss driving this out-flow. \\

\textbf{HD 86226 c}: A recent discovery around a nearby, bright G1V star with V = 7.93 mag, HD 86226 c is another excellent target for observing atmospheric escape \citep[][]{teske2020a}. As shown in \Cref{fig:HD 86226 c}, our core-powered mass-loss model predicts mass-loss rates between 6 $\times 10^4$ g/s to 6 $\times 10^9$ g/s (95\% confidence interval).\\

\textbf{Kepler-60 d, Kepler-60 b, EPIC 249893012 b, Kepler-107 c, Kepler-80 e, and Kepler-138 d}: At the time of writing, to the best of our knowledge, there are no published studies that perform theoretical or observational analysis to determine atmospheric mass-loss from these planets. As shown in \Cref{fig:losing_atm} and \Cref{table:losing_atm}, our results predict that these planets are likely undergoing considerable atmospheric mass-loss. Host stars of these planets are unfortunately dimmer: EPIC 249893012 with V = 11.4 mag \citep[][]{hidalgo2020a}, Kepler-107 c \citep[][]{rowe2014a}, Kepler-138 with V = 12.9 mag \citep[][]{rowe2014a}, Kepler-60 with V = 14.5 mag \citep[][]{steffen2013a,hadden2014a} and Kepler-80 with V = 15.0 mag \citep[][]{muirhead2012a}. Nonetheless, for completeness, we give the mass-loss rates of this set of planets in \Cref{table:losing_atm} and \Cref{fig:losing_atm} with the hope that their  characterization may come within reach one day with future facilities.

\subsubsection{Sensitivity of mass-loss results to the choice of the exact $\rho_{c\ast}$ prior}
To estimate the posterior distributions of planet and stellar properties that characterize a planet's evolution and mass-loss, we assumed that the prior for $\rho_{c\ast}$ is a Gaussian distribution with a mean $\mu_{\rho_{c\ast}}=$ 5 g/cm$^3$ and a standard deviation $\sigma_{\rho_{c\ast}} = 1$ g/cm$^3$. This assumption is motivated by previous observational and theoretical studies \citep[e.g.][]{bower2019a,doyle2019a,gupta2020a,rogersj2020a}. Furthermore, our Bayesian inference model lets us account for the modelling degeneracy between the density of a planet's core and (non-)existence of a hydrogen atmosphere \citep[e.g.][]{rogers2010a,zeng2019a,mousis2020a}. Nevertheless, to check if our best planetary candidates for mass-loss are sensitive to the exact choice of prior for $\rho_{c\ast}$, we repeated our analysis assuming that this prior is a Gaussian distribution with $\mu_{\rho_{c\ast}}=$ 4, 5 and 5.5 g/cm$^3$ and with standard deviations $\sigma_{\rho_{c\ast}}$ equal to 1 and 2 g/cm$^3$. We find that except for Kepler-60 b, all the listed planets yield mass-loss rates similar to those reported in \Cref{table:losing_atm}. For Kepler-60 b, when $\mu_{\rho_{c\ast}} =$ 4 g/cm$^3$, the posterior for $\rho_{c\ast}$ indicates that the planet is more likely to have a significant ice-fraction by mass and/or a secondary atmosphere abundant with high molecular weight gases. Planets that are highly likely to fall into this category are discussed in the next section.

\subsubsection{Consequence of mass-fractionation on estimating mass-loss rates}

In this work, while we do not explicitly account for any mass-fractionation during atmospheric escape, we did repeat our analysis to gauge its impact on the evolution of our top planet candidates. Photoevaporation studies have shown that mass-fractionation can be significant for planets that have been losing their atmospheres for several Gyr \citep[e.g.][]{hu2015a,malsky2020a}. Using the mass-fractionation prescription described in \citet[][]{biersteker2021a}, we find that, for our top planet candidates, the mean molecular weight of the atmospheres only increases by $\sim 0.1$ to $1\%$ at 97.5$^{th}$-percentile by the end of their evolution. The reason for this small increase comes from the fact that these planets have not yet lost enough atmospheric mass at sufficiently low rates to lead to significantly fractionated residual envelopes.
These results thus suggest that for the planets in question, mass-fractionation has likely not played a crucial role in their evolution to date. Furthermore, while we do find that these planets should be experiencing considerable mass-fractionation today - leading to an escape flux of heavier species that is lower by $\sim10\%$ in comparison to the case with no mass-fractionation - it is reasonable to neglect this because a 10\% decrease in escape flux, even with a bulk mass-loss rate of $10^9$ g/s, is well within measurement uncertainties.

Furthermore, in our simulations, we did not account for any opacity changes with composition. As a planet gets more enriched in metals due to mass fractionation, its atmospheric opacity should increase. This will make it harder for the planet to cool and it will thus stay inflated for a longer time. As apparent from \Cref{eq:M_loss_rate_B}, a larger radiative-convective boundary will result in a higher mass-loss rate which, in turn, will lead to lower mass-fractionation \citep[e.g.][]{zahnle1990a}. This suggests that any increase in atmospheric opacity due to mass-fractionation should simply slow down the rate of fractionation itself. Therefore, our estimate of a $\sim 0.1$ to $1\%$ (97.5$^{th}$-percentile) increase in the mean molecular weight of the atmosphere could in reality be even lower and accounting for mass-fractionation in the atmosphere could, in fact, lead to even higher mass-loss rates at later times. We thus conclude that it is reasonable to neglect mass-fractionation for the purpose of this work.

\begin{table*}
\centering
{
\renewcommand{\arraystretch}{2}
\begin{tabular}{c  c  c  c  c  c  c  c  c} 
\hline
Planet & $\tau_\ast$ (Gyr) & $M_{\ast}$ ($M_{\odot}$) & $S_{p}$ ($S_{\oplus}$) & $R_p$ ($R_{\oplus}$) & $M_p$ ($M_{\oplus}$) & References \\ 
\hline

WASP-47 e & $6.5^{+2.6}_{-1.2}$ & $1.04^{+0.03}_{-0.03}$ & $3832.12_{+237.39}^{-237.39}$ & $1.81^{+0.03}_{-0.03}$ & $6.91^{+0.81}_{-0.83}$ & \citet{dai2019a}\\

Kepler-78 b & $0.63^{+0.15}_{-0.15}$ & $0.779^{+0.03}_{-0.05}$ & $4220.00_{+220.30}^{-220.30}$ & $1.23^{+0.02}_{-0.02}$ & $1.77^{+0.24}_{-0.25}$  & \citet{dai2019a}\\ 

Kepler-10 b & $8.52^{+3.59}_{-3.52}$ & $0.93^{+0.07}_{-0.06}$ & $3849.06_{+314.55}^{-292.28}$ & $1.46^{+0.04}_{-0.03}$ & $3.57^{+0.51}_{-0.53}$ & \citet{berger2020a,berger2020b}\\

CoRoT-7 b & $1.32^{+0.75}_{-0.75}$ & $0.88^{+0.03}_{-0.03}$ & $1757.21_{+111.35}^{-111.35}$ & $1.58^{+0.1}_{-0.1}$ & $4.6^{+1.1}_{-1.2}$ & \citet{dai2019a}\\

HD 80653 b & $2.67^{+1.2}_{-1.2}$ & $1.18^{+0.04}_{-0.04}$ & $6295.58_{+322.86}^{-322.86}$ & $1.61^{+0.07}_{-0.07}$ & $5.6^{+0.43}_{-0.43}$  & \citet{frustagli2020a}\\

55 Cnc e & $10.2^{+2.5}_{-2.5}$ & $0.87^{+0.05}_{-0.05}$ & $2504.99_{+128.17}^{-128.17}$ & $1.90^{+0.04}_{-0.04}$ & $7.74^{+0.37}_{-0.3}$  & \citet{dai2019a}\\

Kepler-36 b & $7.78^{+1.33}_{-1.38}$ & $1.03^{+0.07}_{-0.06}$ & $218.78_{+18.61}^{-16.98}$ & $1.52^{+0.06}_{-0.05}$ & $3.83^{+0.11}_{-0.10} $ & \citet{berger2020a,berger2020b}\\

\hline

\end{tabular}
\caption{List of planets that are likely to have secondary atmospheres abundant with heavy molecular weight volatiles, interiors with significant ice-fractions, or both, and their observed parameters (host star's age ($\tau_\ast$) and mass (M$_\ast$), planet's insolation flux (S$_p$), radius (R$_p$), mass (M$_p$) and density when scaled to an Earth mass ($\rho_{p\ast}$)). Values are reported with 1 $\sigma$ errors. The last column lists the references from where all the planet and host-star properties are taken from, unless noted otherwise below. Mass estimates for Kepler 10 b and Kepler 36-b were taken from \citet{dai2019a} and \citet{vissapragada2020a}, respectively, whereas age estimates for WASP-47 e and Kepler-78 b were taken from \citet{almenara2016a} and \citet{howard2013a}, respectively.
}
\label{table:jupi}
}
\end{table*}

\subsection{Planet candidates that are best modeled with secondary atmospheres, low density interiors, or both}\label{sec:super-earths}

Among the 78 planets we analyze, we find that several could not have retained their primordial hydrogen atmospheres to this day. Among these super-Earths, we find that a few of them have observed bulk densities that are much lower than that of the Earth. These relatively low-density super-Earths include 55 Cnc e, WASP-47 e, Kepler-78 b, Kepler-10 b, CoRoT-7 b, HD 80653 b, and Kepler-36 b (see Table \ref{table:jupi}). There is, however, a very low but non-zero likelihood that two of these planets, 55 Cnc e and Kepler-36 b, have indeed hydrogen atmospheres and are {losing} it rapidly today. All of them, except Kepler-36 b, have ultra-short orbital periods and experience high insolation flux (see \Cref{fig:planet_locns}).

\begin{figure*}
\centering
\includegraphics[width=0.75\textwidth,trim=0 0 0 0,clip]{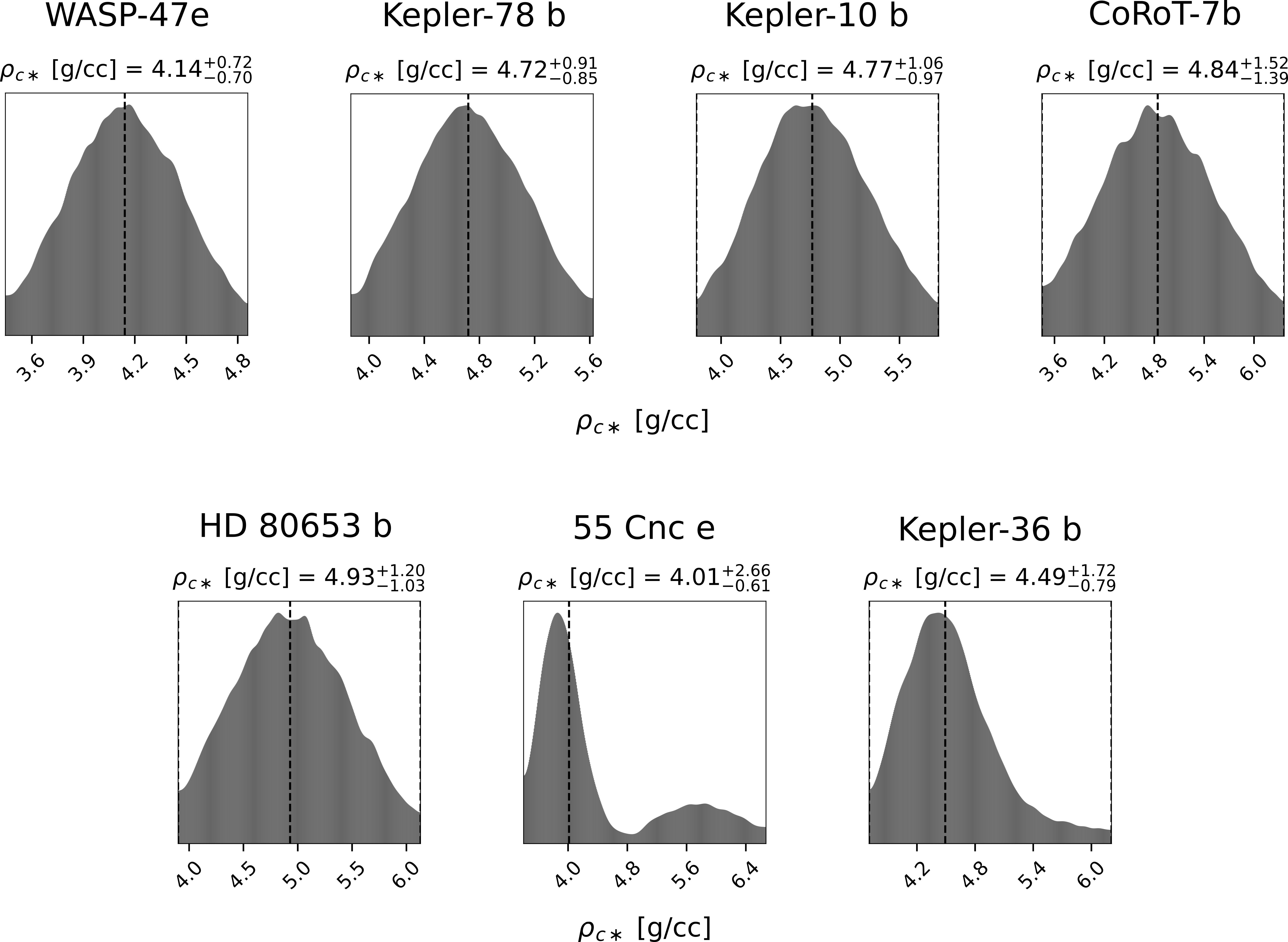} 

\caption{Posterior distributions of $\rho_{c\ast}$ for planets that are likely to have lost their primordial hydrogen atmospheres by today and have bulk densities $\rho_{c\ast} \lesssim 5$ g/cm$^3$. Values are reported with 2 $\sigma$ errors and vertical dashed lines denote the median values. Since these planets have bulk densities much lower than that for Earth, it is likely that they either have secondary atmospheres abundant with heavy molecular weight volatiles, interiors with significant ice-fractions by mass, or both. In addition, we find that 55 Cnc e and Kepler-36 b could also be consistent with having a hydrogen-dominated atmosphere today with a probability of 26\% and 10\%, respectively.}
\label{fig:super_earths}
\end{figure*}

\Cref{fig:super_earths} shows the posterior distributions for $\rho_{c\ast}$ for the planets in Table \ref{table:jupi}. Since our analysis shows that there is a very low likelihood that any of these planets are left with their primordial hydrogen atmospheres today, these plots simply tell us what the bulk compositions of the respective planets could be. It is then surprising that a subset of super-Earths, which are the focus of this sub-section, have low bulk densities of $\rho_{c\ast}$ $\lesssim$ 5 g/cm$^3$. A possible explanation for this is that the planets listed in Table \ref{table:jupi} harbor a secondary atmosphere dominated by high molecular weight species, an interior with significant ice-fraction by mass ($\gtrsim$ 10-20\%), or both.

A source for a high mean molecular weight atmosphere could be outgassing from a planet's interior/magma oceans \citep[e.g.][]{elkins-tanton2008a,bower2019a,chachan2018a,kite2020a} or delivery of volatiles by impactors such as comets { \citep[e.g.][]{raymond2004a,Schlichting2015,kral2020a}}. Once a planet has such a high molecular weight secondary atmosphere, it is typically not possible to lose it through hydrodynamic escape.

On the other hand, a significant amount of ice in a planet's core could also explain a lower $\rho_{c\ast}$, even if the planet has lost all its primordial atmosphere \citep[e.g.][]{rogers2010a,dorn2017a,zeng2019a,mousis2020a}. Such planets would likely have formed and migrated from outside the ice-line. As shown in \Cref{fig:planet_locns}, most of these planets are on ultra-short orbital periods and experience $\sim$ 1000 times more insolation than Earth. It is thus likely that these planets also have steam atmospheres \citep[e.g.][]{mousis2020a,nadejda2020a}. Furthermore, studies have also hypothesized that low bulk densities could be an artifact of planets having exotic interior compositions, for instance, certain super-Earths might not have a core and could instead be enriched in Ca and Al minerals \citep[][]{dorn2017b}. Specifically targeting these planets in future surveys could help us to better understand the wide-variety of possible super-Earths and to constrain the role of migration in sculpting the planetary architectures we observe today.

Among the listed planets, 55 Cnc e, WASP-47 e, Kepler-78 b, Kepler-10 b, CoRot-7 b and Kepler 36-b have been a focus of several theoretical studies. However, to the best of our knowledge, attempts have only been published to detect and characterize an envelope around 55 Cnc e. Below, we discuss these past studies in the context of this work.

\textbf{55 Cnc e}: Among the planets that likely lost their primordial atmospheres, and even otherwise, 55 Cnc e is perhaps the most studied small exoplanet to date. It was first detected almost two decades ago by \citet{mcarthur2004a} and because it orbits a bright Sun-like star with V = 5.95 mag, it has not just been subject of theoretical work but also of numerous observations.

While multiple studies have reported a detection of an atmosphere of 55 Cnc e \citep[e.g.][]{ridden-harper2016a,tsiaras2016a,demory2016a}, their interpretation of the atmosphere's composition differ significantly \citep[e.g.][]{ehrenreich2012a,tsiaras2016a,esteves2017a,zhang2020a,tabernero2020a}. For instance, \citet{ehrenreich2012a} and \citet{zhang2020a} have reported non-detections of hydrogen Ly$\alpha$ and helium 1083 nm triplet absorptions, respectively. These studies thus conclude that it is likely that the planet's atmosphere is not hydrogen-dominated today. This however doesn't preclude that the atmosphere might have a mass-loss rate lower than what can be detected, which is estimated by the authors to be $\sim 3 \times 10^8 - 10^9$ g/s. Alternatively, the atmosphere might have high latitude clouds. Using infrared phase curves collected by Spitzer, \citet[][]{demory2016a} found a large day-night temperature contrast and an off-set of the hottest spot towards east of the sub-stellar point. Based on analysis of such observations, \citet[][]{demory2016a} and \citet{angelo2017a} have argued that 55 Cnc e should have a substantial atmosphere of heavy molecular weight species. \citet{angelo2017a} further argued that these Spitzer phase curve observations can't be explained if the planet had a magma ocean or a tenuous mineral atmosphere because heat cannot be transferred fast enough \citep[e.g.][]{kite2016a}. In addition, \citet[][]{tsiaras2016a}, \citet[][]{esteves2017a} and \citet[][]{jindal2020a} have ruled out the presence of a water dominated atmosphere - based on their analysis of high resolution spectral data collected for the planet. There is also the inexplicable variability in thermal emission from 55 Cnc e which is hard to explain with a simple secondary atmosphere model \citep[][]{demory2016b}. On the other hand, \citet{tsiaras2016a} reported strong absorption features at 1.42 and 1.54 $\mu$m. HCN is identified as the cause of these absorption features and their study finds that these features are strongly suggestive of a hydrogen rich atmosphere.

Our results, unfortunately, do not help in resolving the debate over a hydrogen dominated v.s. secondary atmosphere of 55 Cnc e. Specifically, we find a 74\% probability that $\rho_c* < 4.8$~g/cc (i.e., solutions that require a low-density interior and/or secondary atmosphere) and a 26\% probability that 55 Cnc e still has a hydrogen atmosphere (see bimodality in the $\rho_{c\ast}$ posterior distribution in \Cref{fig:super_earths}). If the latter is true, 55 Cnc e should be losing its atmosphere at a considerable rate. While the 50$^{th}$-percentile or median mass-loss rate for 55 Cnc e is $\sim 0$ g/s, the 84$^{th}$-percentile and 97.5$^{th}$-percentile mass-loss rates are $5 \times 10^8$ g/s and $1 \times 10^9$ g/s, respectively. Interestingly, these estimates are comparable to the threshold mass-loss rates for detection of an escaping atmosphere from 55 Cnc e \citep{ehrenreich2012a,zhang2020a}.\\

\textbf{WASP-47 e, Kepler-78 b, Kepler-10 b, CoRot-7 b, and Kepler-36 b}: These planets have been the focus of several theoretical studies that conjecture - in agreement with our analysis - that they should either have secondary atmospheres (and magma oceans) or/and low density interiors \citep[e.g.][]{schaefer2009a,ito2015a,dorn2017b,dai2019a,chao2020a}. Unfortunately, their host stars are much dimmer than 55 Cnc with V $\sim$ 11 to 12 mag \citep[][]{leger2009a,queloz2009a,howard2013a,becker2015a,batalha2011a}.

There is more to Kepler-36 b, however, as it is also compatible with having a hydrogen atmosphere today. Although this is very unlikely according to our analysis, Kepler-36 b could have a mass-loss rate of $3.5 \times 10^8$ g/s at 97.5$^{th}$ percentile. Kepler-36 b is also peculiar in receiving an order-of magnitude lower insolation flux than all the other planets that likely do not have a primordial hydrogen atmosphere today.\\

\textbf{HD 80653 b}: Recently discovered by \citet{frustagli2020a}, HD 80653 b is a small planet around a relatively bright G2 star with V = 9.5 mag. Frustagli et al. speculate that the planet is likely to have lost its hydrogen atmosphere because it is highly irradiated - more than 6000 times than Earth. Our results show that this is indeed the case if the planet's evolution was dominated by core-powered mass-loss. In addition, it has a low bulk density for its size and mass. These factors make it yet another interesting target for further investigations.

\section{Conclusion} \label{sec:conclusion}

In this paper, we investigate which planets could be undergoing considerable atmospheric mass-loss today if their evolution was dominated by core-powered mass-loss. Of the 78 planets in our sample, we find ten planets with median mass-loss rates of $\gtrsim 10^7$ g/s: pi Men c, Kepler-60 d, Kepler-60 b, HD 86226 c, EPIC 249893012 b, Kepler-107 c, HD 219134 b, Kepler-80 e, Kepler-138 d and GJ 9827 d. 

The covariance plots for these targets, unsurprisingly, show correlations between different planetary parameters. Notably, we find that the estimated mass-loss rates are especially sensitive to the uncertainties in a planet's mass measurements. This is especially apparent for planets such as HD 86226c and GJ 9827 d for which our mass-loss estimates vary over $\sim$ 5-10 orders of magnitude for planet masses within 2$\sigma$ of the median. Furthermore, our results show a two-peaked posterior mass-loss distribution for several planets - peaking at $\sim 0$ g/s and at $\gtrsim 10^7$ g/s. This indicates that a number of planets from this category, such as HD 219134 b and EPIC 249893012 b, have a low - but non-zero - probability of not losing any atmosphere at all and could rather have a secondary atmosphere and/or low density interiors. While the likelihood of this is typically lower than 10\%, Kepler-60 d and Kepler-60 b are only 2 and 1.5 times more likely to have hydrogen-dominated atmospheres, respectively.

Among these planets, HD 219134 b is a especially interesting target because while our analysis suggests that it is likely to have a hydrogen dominated atmosphere, photoevaporation studies predict the opposite \citep[][]{dorn2018a,kubyshkina2018a}. Detecting atmospheric mass-loss from HD 219134 b could therefore directly confirm that core-powered mass-loss is driving its evolution.

Finally, even though all the planets in this category are likely undergoing fairly rapid atmospheric escape, their 95$^{th}$-percentile mass-loss rates are still lower than $\sim 9 \times 10^{9}$ g/s. These lower mass-loss rates might therefore explain the non-detection of Ly$\alpha$ absorption or the He 1083 nm triplet feature for planets such as pi Men c \citep[][]{garcia2020a} and GJ 9827 b \citep[][]{kasper2020a,carleo2021a}.

In addition, we also identify seven planets for which our core-powered mass-loss model predicts no residual hydrogen envelopes today, and whose mass and radius measurements thus suggest that they likely have have secondary atmospheres and/or low density interiors. The planets in this list are 55 Cnc e, WASP-47 e, Kepler-78 b, Kepler-10 b, CoRoT-7 b, HD 80653 b, and Kepler-36 b. Among these, our results are somewhat ambiguous for 55 Cnc e and Kepler-36 b because they are also compatible, at a lower probability, with having hydrogen-dominated atmospheres. For instance, 55 Cnc e is only about three times more likely to have a secondary atmosphere and/or low density interior than having a hydrogen dominated envelope according to our analysis.

\subsection{Future Work}
Studies on core-powered mass-loss and photoevaporation have shown that both mechanisms can, independently, explain a multitude of observations associated with the observed radius valley \citep[e.g.][]{owen2017a,ginzburg2018a,wu2019a,gupta2019a,gupta2020a,rogersj2020a}. We assume in this paper that core-powered mass-loss dominates planet evolution. It is however likely that both mechanisms take place simultaneously to a certain degree. Moreover, it is possible that the two processes dictate the atmospheric mass-loss to varying degrees in different parts of the parameter space of the relevant planet and stellar properties. In future studies, we therefore plan to tackle this by combining the two mechanisms to develop a more comprehensive understanding of how planet atmospheres evolve.

\subsection{Outlook}
In this work, we have identified ten planets that could be undergoing atmospheric mass-loss today, as well as, a list of seven planets that likely have secondary atmospheres, ice rich/iron poor interiors, or both. We hope that the results from this work will aid in the target selection process for future observations. As alluded to earlier, remarkable efforts are continually being made in this direction \citep[e.g.][]{ehrenreich2012a,demory2016a,ridden-harper2016a,benneke2019a,vissapragada2020a,kasper2020a,zhang2020a}. The larger sensitivity and spectral range offered by the \textit{James Webb Space Telescope} \citep[e.g.][]{greene2016a,allart2018a}, scheduled for a launch later this year, and the next-generation ground-based telescopes \citep[e.g.][]{hood2020a,vissapragada2020a} should provide us with new data and new insights in the near future.

\section*{Acknowledgements}
{We thank the anonymous referee for valuable comments that helped improve the manuscript.} A.G. and H.E.S. acknowledge support from the Future Investigators in NASA Earth and Space Science and Technology (FINESST) grant 80NSSC20K1372. H.E.S. further acknowledges support from NASA under grant number $17-\rm{XRP}17\_2-0055$ issued through the Exoplanet Research Program. 


\section*{Data Availability}
The exoplanet data used here is summarized in \Cref{table:sim_char,table:jupi} and further details can be found in the relevant references given in this paper. For data underlying the generated figures, please contact the corresponding author.

\bibliographystyle{mnras}
\bibliography{planet_evo}

\bsp	
\label{lastpage}
\end{document}